\documentclass[conference]{IEEEtran}
\IEEEoverridecommandlockouts
\usepackage{cite}
\usepackage{amsmath,amssymb,amsfonts}
\usepackage{graphicx}
\usepackage{textcomp}
\usepackage{booktabs}
\usepackage{xcolor}
\usepackage{subfigure}
\usepackage[T1]{fontenc}
\usepackage{url}
\def\BibTeX{{\rm B\kern-.05em{\sc i\kern-.025em b}\kern-.08em
    T\kern-.1667em\lower.7ex\hbox{E}\kern-.125emX}}
    
\begin{document}

\title{SMOTEC: An Edge Computing Testbed for Adaptive Smart Mobility Experimentation\\
%\thanks{This work is funded by Alan Turing??}
}

\author{\IEEEauthorblockN{Zeinab Nezami}
\IEEEauthorblockA{\textit{School of Computing} \\
\textit{University of Leeds}\\
Leeds, United Kingdom \\
z.nezami@leeds.ac.uk}
\and
\IEEEauthorblockN{Evangelos Pournaras}
\IEEEauthorblockA{\textit{School of Computing} \\
\textit{University of Leeds}\\
Leeds, United Kingdom \\
e.pournaras@leeds.ac.uk}
\and
\IEEEauthorblockN{Amir Borzouie}
\IEEEauthorblockA{\textit{School of Computing} \\
\textit{University of Leeds}\\
Leeds, United Kingdom \\
sc22a2b@leeds.ac.uk}
\and
\IEEEauthorblockN{Jie Xu}
\IEEEauthorblockA{\textit{School of Computing} \\
\textit{University of Leeds}\\
Leeds, United Kingdom\\
j.xu@leeds.ac.uk}

}

\maketitle

\begin{abstract}

Smart mobility becomes paramount for meeting net-zero targets. However, autonomous, self-driving and electric vehicles require more than ever before an efficient, resilient and trustworthy computational offloading backbone that expands throughout the edge-to-cloud continuum. Utilizing on-demand heterogeneous computational resources for smart mobility is challenging and often cost-ineffective. This paper introduces SMOTEC, a novel open-source testbed for adaptive smart mobility experimentation with edge computing. SMOTEC provides for the first time a modular end-to-end instrumentation for prototyping and optimizing placement of intelligence services on edge devices such as augmented reality and real-time traffic monitoring. SMOTEC supports a plug-and-play Docker container integration of the SUMO simulator for urban mobility, Raspberry Pi edge devices communicating via ZeroMQ and EPOS for an AI-based decentralized load balancing across edge-to-cloud. All components are orchestrated by the K3s lightweight Kubernetes. A proof-of-concept of self-optimized service placements for traffic monitoring from Munich demonstrates in practice the applicability and cost-effectiveness of SMOTEC.
%Edge computing is one of the principal supporters for intelligent transportation system and autonomous vehicles to improve safety, enhance efficiency, and decrease traffic congestion. While current evaluations are more based on simulators and their results fidelity is under debate, it is essential to test services and performance of applications in this domain before they are deployed to the production environment.
%This paper presents SMOTEC, an edge computing testbed built with real devices and open sources to support smart mobility experimentation. SMOTEC supports mobile end-devices using mobility and migration management strategies and simple application programming interfaces. Edge workloads are run as sets of containers with access to compute resources on an expandable cluster of light-weight edge nodes to decrease implementation costs. 
%While most of the related works leave the offloading decision to the user, our proposed service distributor module automatically determines which edge nodes to place Docker containers on according to service requirements specified by the testbed users. Using a simple proof of concept traffic monitoring application, we demonstrate that SMOTEC is easy to configure, scalable, and provides users with a rich overall view of their resource utilization.
\end{abstract}


\section{Introduction}

The penetration of smart mobility, including autonomous, self-driving and electric vehicles, is transforming cities, providing new opportunities for more efficient and sustainable transport. From such penetration, socio-technical infrastructures emerge as complex and interdependent: connected vehicles generating novel traffic flows require coordinated mobility services to meet safeguards and net-zero goals, while generating massive privacy-sensitive (training) data. These require in turn real-time processing by resource-intensive artificial intelligence (AI) algorithms. Via flexible utilization of computational resources, the edge-to-cloud continuum turns out to be an enabler paradigm for the smart mobility niche. So far, the research community lacks of general-purpose instruments for low-cost and low-complexity prototyping, deployments and experimentation of smart mobility solutions based on edge computing~\cite{berman2014geni,ertin2006kansei,keahey2020lessons}. But also the edge computing research community is over-relying on synthetic data and simulation tools of limited realism~\cite{carvalho2022can,svorobej2019simulating}.

This is a research gap that this paper bridges via a new testbed: \emph{Smart Mobility Services On The Edge Computing (SMOTEC)}. SMOTEC is designed to support scientists of different disciplines (computing, transport, social and environmental science) to study problems that interlink smart mobility and computation. This includes the impact of smart mobility applications (e.g. self-driving operations, augmented reality, traffic monitoring and control, multi-modal transportation) on edge-to-cloud computing infrastructures (e.g. optimized resource allocation, load balancing) and vice versa. SMOTEC maps spatio-temporal service requests of mobile agents (vehicles, pedestrians, cyclists, drones, etc.) to heterogeneous computing resources within the edge-to-cloud continuum, while possessing autonomic capabilities to improve quality of service (QoS) via self-adaptive service placements~\cite{Nezami2021}. 

SMOTEC is designed on generic and extensible Application Programming Interfaces (APIs), whose implementation integrates the SUMO urban mobility simulator~\cite{behrisch2011sumo}, interconnected Raspberry Pi devices (referred to in the rest of this paper as Pis) for edge computation~\cite{gizinski2022design} and the multi-agent collective learning approach of EPOS~\cite{pournaras2018decentralized,pournaras2020collective,Nezami2021} for load balancing and optimizing service placements within the edge infrastructure. All these modules can be easily replaced by other ones to meet the needs of different application and experimental scenarios. Users' configurations are quick and simple by running code in containers without incurring the overhead of virtualization or security infrastructure of bare-metal privileged access. A distributed load-balancing scenario of service placement for monitoring traffic flows in Munich city is demonstrated as a proof-of-concept. 

The contributions of this work are outlined as follows: (i) A general-purpose and modular testbed model for adaptive smart mobility experimentation based on edge computing infrastructure. (ii) An instantiation of the testbed model via the plug-and-play integration of the SUMO simulator, the interactive Pis and the EPOS learning algorithm for service placements. (iii) An open-source software artifact\footnote{\url{https://doi.org/10.5281/zenodo.8167871}} \footnote{ \url{https://github.com/DISC-Systems-Lab/SMOTEC}} to encourage further work, applicability and adoption of the proposed testbed. (iv) Insights from the applicability of SMOTEC to a load balancing scenario for traffic monitoring in Munich. 

This paper is organized as follows: Section~\ref{sec:related-work} compares SMOTEC with related work. Section~\ref{sec:model} illustrates the testbed design model. Section~\ref{sec:implementation} introduces the implemented testbed architecture of SMOTEC. Section~\ref{sec:workflow} introduces the workflow of a smart mobility application scenario. Section~\ref{sec:evaluation} illustrates a proof-of-concept evaluation of SMOTEC. Section~\ref{sec:conclusion} concludes this paper and outlines future work.

\section{Related work}\label{sec:related-work}

SMOTEC is distinguished from existing edge computing testbeds~\cite{vasisht2017farmbeats,zhang2019hetero,gedawy2016cumulus,hao2018edge,meng2019dedas,munoz2017adrenaline,pan2016homecloud} in terms of its capabilities, generality/abstractions and applicability. Most are developed to support customized use cases, e.g., benchmarking artificial intelligence algorithms~\cite{hao2018edge,zhang2019hetero}, offloading approaches~\cite{gedawy2016cumulus}, hardware paradigms~\cite{pan2016homecloud,royuela2022testbed}, custom workloads~\cite{meng2019dedas,munoz2017adrenaline,vasisht2017farmbeats}, or edge networking~\cite{rimal2018experimental,meng2019dedas,munoz2017adrenaline}. 
These testbeds come with a limited scope, providing restricted deployments, openness, modularity/configurability and do not easily scale. 

\par Xu \textit{et al.}~\cite{xu2020support} propose a fog computing testbed, piFogBedII, to support testing of mobile fog applications. PiFogBedII is an enhancement of PiFogBed~\cite{xu2019pifogbed}, built with Pis by adding mobility and migration management strategies. PROWESS~\cite{boubin2022prowess} is a general-purpose edge computing testbed for evaluating resource constrained applications as a set of containers. Fogbed~\cite{coutinho2018fogbed} leverages Docker containers and a mininet virtual network to support adding and removing containers in the network topology at any time during experimentation. EdgeNet~\cite{cappos2018edgenet} is a kubernetes cluster that consists of a master node which manages a set of globally distributed worker nodes. 
The main limitation of these works~\cite{coutinho2018fogbed,cappos2018edgenet,rimal2018experimental,xu2020support,xu2019pifogbed,boubin2022prowess} is a lack of online adaptation to the environment settings, meaning that the resource allocation/migration decisions require manual offline interventions by users.
%In the continuously changing environment of edge computing, such decisions need to dynamically take account of new patterns in resources and mobility of end-devices to meet the QoS requirements at scale.

\par TRAPP~\cite{Gerostathopoulos2019} is a testbed for smart mobility services~\cite{Davis2021} that brings together transport domain knowledge (SUMO traffic microscopic simulator) and domain-independent multi-agent distributed intelligence (collective learning with EPOS~\cite{pournaras2018decentralized,pournaras2020collective}). This provides significant flexibility for online adaptations in distributed optimization of traffic flows~\cite{Davis2021}. This paper makes a significant advancement to TRAPP by expanding its scope to edge computing and allowing for the first time to experiment and study in an integrated way the full continuum of mobility-to-edge-to-cloud infrastructures. SMOTEC leverages SUMO as its mobility service provider. It can be replaced with a real-time mobility service or spatial and transport modeling tools such as Harmony~\cite{yfantis2021software}.

\par In overall, the proposed testbed is a significant advancement in the field by offering a remarkable level of online, automated and scalable resource adaptations: service placements are autonomously managed during runtime using decentralized intelligence, while providing high realism of smart mobility via a plug and play integration of SUMO or other simulation modules and real-time mobility monitoring services. 

\section{Testbed Design Model}\label{sec:model}

The proposed testbed supports researchers and developers to prototype and test edge computing services with simulated and real-world smart mobility workloads. As shown in Figure~\ref{fig:arch}, SMOTEC is by design a distributed system that consists of at least one system orchestrator
and a set of edge nodes. Although an orchestrator is basically an edge node with more functionalities to coordinate all the edge infrastructure in the cluster, the orchestrator and edge nodes can be separate entities and part of a geographically distributed pool of
interconnected computational resources. The main testbed modules are outlined in the rest of this section.

\noindent \textbf{Edge nodes}: The edge infrastructure offers distributed storage/processing capabilities, these are edge nodes in the vicinity of end-users to run smart mobility services with real-time low-latency requirements. An edge node refers to the combination of a base station (or access point) and its co-located edge servers in mobile radio networks. The testbed supports for heterogeneous edge nodes, from well-provisioned and centralized servers to far-flung and lightly provisioned embedded devices.

\noindent \textbf{Agents}: SMOTEC introduces two types of agents: mobile agent and edge agent. A mobile agent is an abstraction of a mobile device (end-users) such as (autonomous) vehicle equipped with a set of sensors (e.g., GPS, camera) for perceiving the surrounding environment (e.g., position, speed) and a set of actuators (e.g., display, acceleration). The mobile agent is in charge of sensing the location of the mobile device and creating/removing communication links with edge nodes according to the mobility profile. Mobility is determined during runtime by the location of the mobile device in relation to the fixed edge nodes. 
An edge agent runs on an edge node and is responsible for responding to the requests received from mobile agents. It decides where to deploy the requested services for the end users to meet QoS requirements and serve its hosted services.

\noindent \textbf{Connector}: Cellular networks have a wide and high-speed communication range, which allows a base station to preserve connectivity and continuously serve a mobile agent (e.g., vehicle). The upcoming 5G/6G cellular networks are leading technologies for native mobile edge computing capabilities~\cite{slawomir2017next,narayanan2022collective} to support mobility applications such as augmented reality with high network capacity and throughput/bandwidth. In SMOTEC, the connector manages the communication network and is responsible for communication between edge agents and service distributor. It connects mobile agents and their serving edge agents together, while synchronizing data pipelines. 

\noindent \textbf{System orchestrator}: The management layer, referred to as orchestrator, maintains a comprehensive view on available resources and running services in the edge network. The testbed API also runs on the orchestrator and enables developers to initiate an experiment with access to SMOTEC resources and APIs. The orchestrator has a small footprint (Minimum: 1 CPU core and 512MB RAM) that can be as light as a Pi. In summary it is responsible for (i) scaling up and down the available resources as required by running applications, (ii) allocating and releasing the storage, networking, and computational resources provided by the service distributor, (iii) storing application images for faster instantiation when required, (iv) providing support for fault and performance monitoring by sharing data about resources and running applications with the monitor module.

\noindent \textbf{Service distributor}: It manages the task of selecting the appropriate hosts for the requested services by mobile agents. The distributor receives edge agent placement selections that takes into consideration service requirements, e.g. CPU and memory, available edge resources, and locations of mobile agents. This module supports user's selection of a service placement strategy or can implement placement policies in the form of Docker container.

\noindent \textbf{Service}: SMOTEC provides smart mobility applications that run in a mobile edge computing environment. A service makes efficient use of the network for the subscribed mobile agents.  

\noindent \textbf{Monitor}: It provides an overall view of resources and their utilization. Information exchanged among agents is recorded in a persistent directory as an experimentation output. 

\section{Testbed Architecture and Implementation}\label{sec:implementation}

\par
\begin{figure}[!htb]
\centering
\includegraphics[clip, trim=5.4cm 14.1cm 3.8cm 8.6cm, width=\columnwidth]{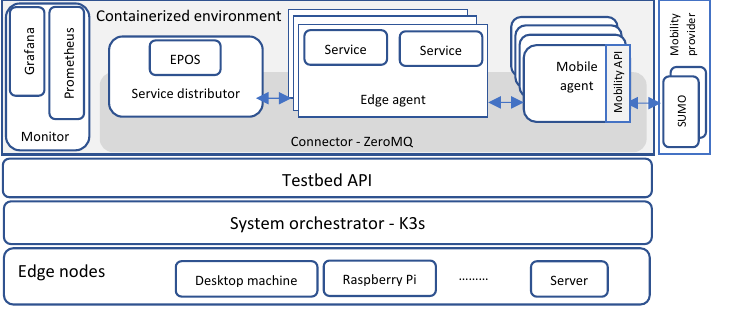}
\caption{The architectural model and implementation of SMOTEC.}
\label{fig:arch}
\end{figure}

 This section introduces the architecture, implementation, and configuration of an easy-to-use edge computing testbed system providing a flexible and modular experimental environment for smart mobility applications via APIs. A 6-node prototype of SMOTEC comprising of one orchestrator and five edge nodes is set up. The testbed provides services to the mobile agents in the form of networked Docker containers hosted on edge nodes and managed by the K3s master node. Deploying services in the form of Docker instances provides a virtualized isolated environment for service execution at each edge node. The general architecture and the essential components of the proposed testbed are illustrated in Figure~\ref{fig:arch}. 

%\fbox{

\begin{figure}[!htb]
\centering
\includegraphics[clip, trim=2.6cm 19cm 8cm 6cm, width=1.0\columnwidth]{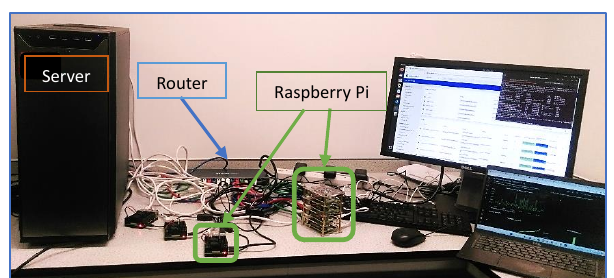}
\caption{The SMOTEC testbed infrastructure.}
\label{fig:testbed}
\end{figure}

\noindent \textbf{Experimentation testbed}: The Pis shown in Figure~\ref{fig:testbed} are used in the implementation and deployment of the testbed: Model 4B with Broadcom BCM2711, Quad core Cortex-A72 64-bit SoC 1.5GHz, 8GB RAM, and 128GB Storage. A TP-Link 16-Port Desktop Gigabit Ethernet Switch is used to connect the edge nodes to the access network via Cat 6 Ethernet cables.

\noindent \textbf{SMOTEC API}: It is the core of the testbed. Users define experiments as (i) a set of edge nodes along with their geographic location and resource specifications of processing power, memory, and storage, (ii) a set of mobile agents (vehicles) with their mobility profile and service requirements such as processing power and memory. The services are provided in the form of Docker containers and instantiated according to their resource requirements for allocations of RAM and CPU. 
SMOTEC follows a modular development; users may test their own smart mobility services in the form of containers. For this, users provide access to their service containers either via Docker Hub or a private Docker repository. 
The testbed API interacts with the orchestrator to instantiate the edge agents and mobile agents on Pis, and then instantiate service containers from pre-created Docker container images. SMOTEC containers are allocated and scheduled using a combination of the K3s scheduler and the service distributor outlined in next subsections.

\noindent \textbf{K3s}: K3s is a lightweight Kubernetes distribution, which is fast to start up and easy to auto-update. It has small memory footprint, simplified configuration and reduced resource usage for resource-constrained IoT and edge environments. The testbed orchestrator, i.e. K3s master, is the module interfacing with the testbed API and managing the edge resource allocation and releasing. Each of the Pis is a K3s worker node enriched with an edge agent for integration with the testbed. 

\noindent \textbf{Raspberry Pi}: SMOTEC is built using low-cost Pis, which can be easily deployed at large-scale. Multiple sensors can be connected to these single board computers to support various smart applications. Each Pi has a fixed location (Cartesian coordinates) provided by developer. Pis are connected to the orchestrator by joining the K3s cluster as worker nodes.

\noindent \textbf{Agents and Docker container}: SMOTEC implements both edge agent and mobile agent in the form of containers. Docker containers are open source, fast with easy startup, lightweight, and preferable for a smooth transition of tested application programs to production environments. Docker containers are also used to deploy the service distributor and users' services on Pis. 

\par SMOTEC instantiates an edge agent on each Pi to make available and manage edge node functionality, from communication to computation. It connects the edge nodes to the connector and guarantees the continuity of the running services via related service migrations. SMOTEC does not make any assumption about edge nodes; the only user information required is the available resources and the location of edge nodes. Such abstraction allows SMOTEC to support the heterogeneous edge resources. This also allows the management of related edge servers and access points as a single entity.
\par
\begin{figure}[!htb]
\centering
\includegraphics[clip, trim=5.5cm 14cm 6cm 9.5cm, width=0.9\columnwidth]{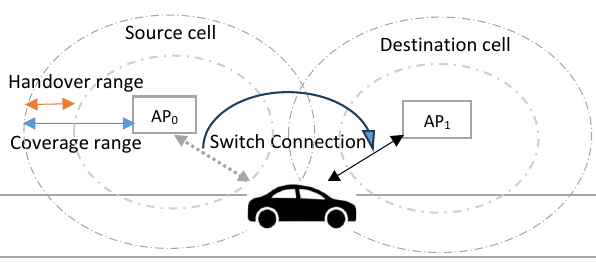}
\caption{Handover procedure during vehicle movement}
\label{fig:mig}
\end{figure}

 A mobile agent container emulates a mobile device and guarantees the security of internal containers. It is in charge of maintaining the agent's connection with the cluster and interacts with its requested service during agent's movement. A mobility module runs on mobile agents to update the mobility profile of the agent and the distance of itself to its current connected edge agent; if the distance is higher than a threshold, it sends a connection (handover) request to another edge agent with the lowest distance, see Figure~\ref{fig:mig}. At the same time the mobile agent sends a notification to its current connected edge agent to inform the agent about the migration.

\par The developed mobility API can be used to implement different mobility models. The evaluation scenario involves a real-world map of Munich city, imported to the SUMO simulator~\cite{behrisch2011sumo} to accurately simulate realistic vehicle movements. The mobility module parses the mobility database information stored in a csv format.
Each csv file represents the mobility profile of a vehicle as a sequence of traversed points during simulation. Each point includes agent's x and y coordinates on the map, its speed in meters per second, the direction in radiant, and the time in which this data is collected.

\noindent \textbf{ZeroMQ}: SMOTEC leverages ZeroMQ as the connector of its components. ZeroMQ is an asynchronous messaging library for edge computing applications. In a real-world scenario, wireless base stations connect edge servers and vehicles. In SMOTEC, edge agents are an abstraction of edge servers and their co-located base station in which ZeroMQ handles the networking part. Here the wireless base stations connect to the wired network nodes, whose location in the network corresponds to wireless base stations in a mobile backhaul network. This emulated infrastructure is compatible with the edge computing model in the 5G network proposed by ETSI~\cite{etsi2018white}. 

\noindent \textbf{EPOS as a decentralized service distributor}: Upon receiving a service request from a mobile agent, SMOTEC automatically determines the Pi that satisfies the service requirements and configures the service Docker container on that Pi. SMOTEC utilizes EPOS~\cite{pournaras2018decentralized,pournaras2020collective}, a decentralized multi-agent system for multi-objective combinatorial optimization, to balance the input workload across the network~\cite{Nezami2021},
while minimizing deadline violations, service deployment cost and services that do not meet hosting requirements. 

\par EPOS performs collective decision-making among edge agents that autonomously generate a set of service placement plans from which they make a choice such that their combination satisfies network-wide (e.g., minimizing over-utilized edge nodes) and individual objectives (e.g., minimizing service execution cost). Then, service containers are scheduled on the selected Pis via the K3s master. The edge agents get informed about the EPOS placement decisions via message passing. EPOS can be replaced with any other coordinator in the form of a Docker image without requiring further reprogramming.

\noindent \textbf{Grafana/Prometheus}: Users can monitor the real-time status of every edge node through the provided open-source interactive data-visualization platform of Grafana\footnote{Grafana. Available at \url{https://grafana.com/} (Accessed 17 July 2023)}, including CPU,memory, and network utilization. Users can customize configured dashboards on Grafana. An output directory that contains the output of the testbed modules is generated.

\section{Testbed workflow}\label{sec:workflow}
\par Assume an application that monitors, in real-time, a number of vehicles passing from different sections of a city. The application consists of two modules: a vehicle flow monitor module referred to as 'collector' running on edge nodes and a client module referred to as 'view' running on end-user devices (e.g., vehicle). As the end-device moves, the view module is connected to the collectors to get an updated local view of traffic; each edge agent periodically exchanges the monitoring data to the collector modules running on its neighbor nodes to provide a shared view of the traffic observed over the city. 

\par
\begin{figure}[!htb]
\centering
\includegraphics[clip, trim=3cm 12.3cm 0.5cm 3.2cm, width=\columnwidth]{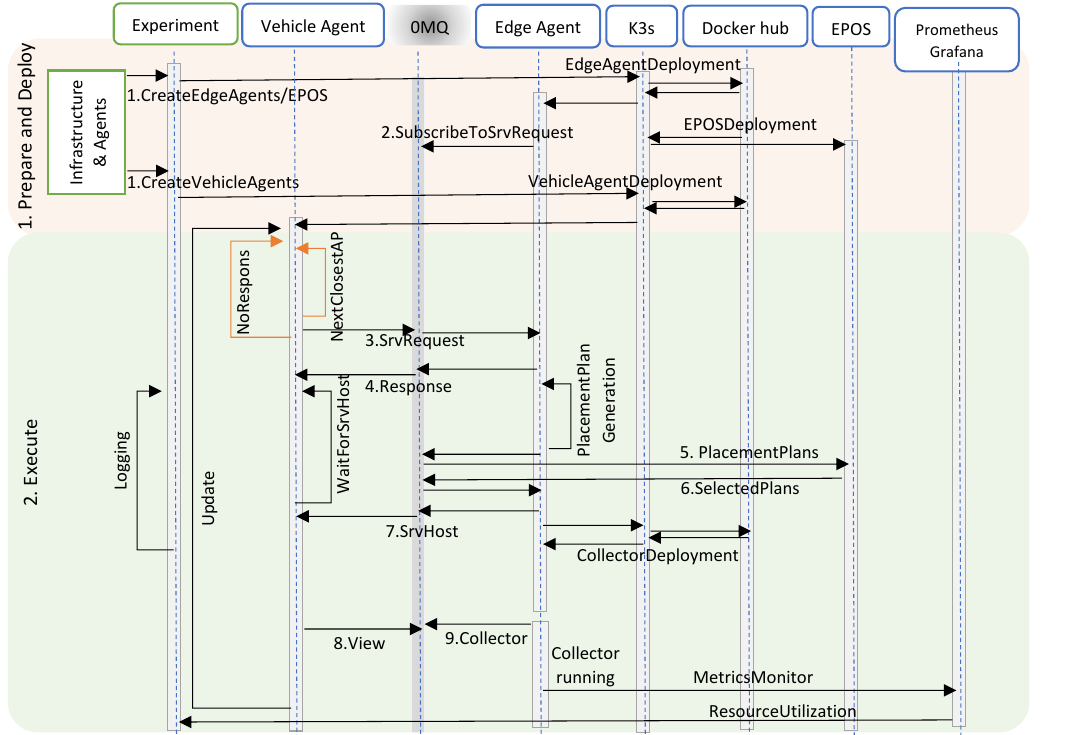}
\caption{Traffic monitoring application workflow}
\label{fig:workflow}
\end{figure}

 The testbed lifecycle includes two stages as shown in Figure~\ref{fig:workflow}: (i) prepare and deploy, (ii) execute. Suppose that the user generates one service Docker image, as the collector, stored on the Docker Hub. The image runs on edge agents. The user utilizes mobile vehicles to continuously monitor the traffic volume of a road over a period of time. For this, at the prepare and deploy stage, the user provides the number of edge nodes and their location, the number of vehicle agents passing through the city of the experiment along with their mobility profiles. This input data and the dataset of mobility profiles are in the form of json and csv files, respectively. 

\par The testbed carries out the experiment by deploying Docker images for every testbed component.
At startup, vehicle agents connect to their closest edge agent in their coverage area to forward their service requests. The edge agent generates a set of service placement plans considering the available edge resources and QoS requirements. Then, the agent coordinates its placement decisions with other edge agents using EPOS to find an system-wide optimized placement. Based on the selected hosts, where services are placed, the edge agent schedules the service deployment of the collector via K3s API. At the same time, the edge agent notifies the vehicle agent about its service host with which the agent can communicate.

\section{Evaluation}\label{sec:evaluation}
\par A proof-of-concept implementation of SMOTEC with the traffic monitoring application of Section~\ref{sec:workflow} are illustrated. For simplicity we run the experiments using three Pis: one orchestrator and two Pis receiving traffic monitoring data from a $7{km^2}$ area of Munich city center. For this test, the service distributor and vehicle agents run on orchestrator. 

\par Experimentation involve 10 vehicles connected to Pis to receive traffic monitoring services. Two scenarios are evaluated: (i) homogeneous, in which all vehicles make the same service requests of CPU (230 MIPS) and memory (350 MB), (ii) heterogeneous, in which all service requests come with different CPU and memory resources, randomly within $[200,300]$ MIPS and $[300,400]$ MB respectively. The CPU and memory capacity of Pis is 2000 MIPS and 4000 MB. EPOS balances the service distribution load by running for 50 learning iterations to minimize the variance of the load over the two Pis.

\par
\begin{figure}[!htb]
    \centering
    \begin{subfigure}{}
        \centering
        \includegraphics[clip, trim=0.0cm 1.3cm 0cm 0cm, width=4cm]{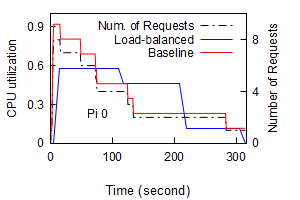}
    \end{subfigure}
    \begin{subfigure}{}
        \centering
        \includegraphics[clip, trim=0.0cm 1.3cm 0cm 0cm, width=4cm]{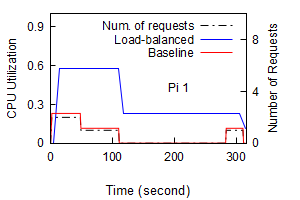}
    \end{subfigure}	
    \centering
    \begin{subfigure}{}
        \centering
        \includegraphics[clip, width=4cm]{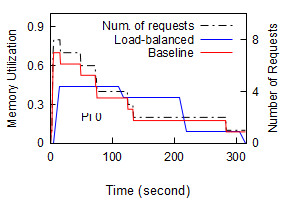}
    \end{subfigure}
    \begin{subfigure}{}
        \centering
        \includegraphics[clip, width=4cm]{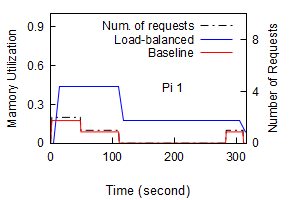}
    \end{subfigure}	
    \caption{Homogeneous scenario: Number of service requests and load distribution for each of the two Pis.}
    \label{fig:homLoad}
\end{figure}

\begin{figure}[!htb]
    \centering
    \begin{subfigure}{}
        \centering
        \includegraphics[clip, width=4cm, trim=0.0cm 1.3cm 0cm 0cm]{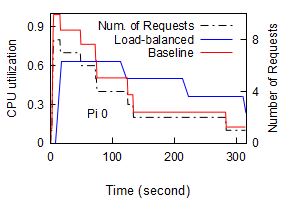}
    \end{subfigure}
    \begin{subfigure}{}
        \centering
        \includegraphics[clip, trim=0.0cm 1.3cm 0cm 0cm, width=4cm]{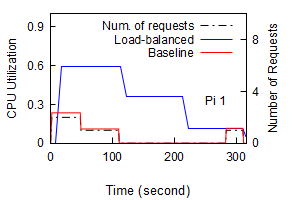}
    \end{subfigure}	
    \centering
    \begin{subfigure}{}
        \centering
        \includegraphics[clip, width=4cm]{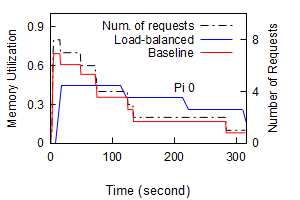}
    \end{subfigure}
    \begin{subfigure}{}
        \centering
         \includegraphics[clip, width=4cm]{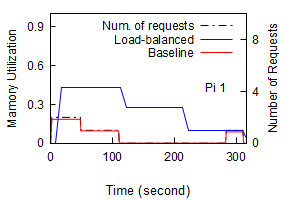}
    \end{subfigure}
    \caption{Heterogeneous scenario: Number of service requests and load distribution for each of the two Pis.}
    \label{fig:hetLoad}
\end{figure}

 The conducted experiments compare a baseline approach (where all received requests from vehicles are served via their connected Pi) with the SMOTEC approach of service distribution with a balanced CPU/Memory utilization of Pis, as shown in Figure~\ref{fig:homLoad} and~\ref{fig:hetLoad}. Each service request originates from one vehicle. The utilization metric shows to what extent Pis are utilized (placed workload/capacity). In the baseline case, the utilization of the one Pi varies from 0.7\% to 99\%, while the other Pi remains under-utilized (<27\% utilization). In the heterogeneous case, SMOTEC shows a CPU and memory utilization for both Pis of up to 62\% and 45\%, and in the homogeneous case, up to 58\% 45\% for both Pis, respectively. This is due to the different placement policy of EPOS in SMOTEC for balancing the load on network resources.

\par The workload balance, measured by the utilization variance among the two Pis, is 0.078 and 0.088 for the homogeneous and heterogeneous scenarios in baseline. Via load balancing, the utilization variance drops to 0.004 and 0.007 respectively. SMOTEC manages to effectively balances by more than 92\% the workload using the integrated EPOS service distributor.

%\par Concerning human resources for testbed system management, the proposed testbed system has no extra maintenance costs in addition to that of the K3s system as the testbed system keeps running without human operations.

\section{Conclusion and Future Work}\label{sec:conclusion}

The proposed testbed provides a significant missing instrumentation for inter-disciplinary research on smart mobility based on edge computing. Experimentation with SMOTEC becomes easier, simpler and less costly. This is because intelligent smart mobility services running on Docker containers can be easily deployed to heterogeneous edge-to-cloud resources via the K3s lightweight Kubernetes, while automatically load balancing service placements in a fully decentralized way. 

SMOTEC APIs generate rich data of high realism for edge computing research that have been so far hard to acquire, evident from simulation approaches dominating related work with low realism. Via a plug-and-play integration of smart mobility modules such as SUMO, complex traffic monitoring scenarios on edge computing can be studied experimentally. This proof-of-concept is demonstrated along with optimized service placements that reduce overloaded edge nodes.

Future work will explore the integration of other smart mobility modules beyond SUMO, and in particular, real-time smart mobility services for augmented reality and autonomous vehicles. We aspire to encourage and advance research on the co-optimization of coupled smart mobility and edge computing infrastructures for meeting net-zero targets. 

\section*{Acknowledgments}
%{\footnotesize
The authors would also like to acknowledge the support of Emmanouil Chaniotakis on the Munich traffic scenario in SUMO. Abhinav Sharma supported the preparation and documentation of the software artifact. The authors would like to thank Satish Kumar and Renyu Yang for their encouragement and support to develop this testbed. This work is supported by an Alan Turing Project. Evangelos Pournaras and Jie Xu are supported by an Alan Turing Fellowship. Evangelos Pournaras is also supported by a UKRI Future Leaders Fellowship (MR\-/W009560\-/1): `\emph{Digitally Assisted Collective Governance of Smart City Commons--ARTIO}' and the European Union, under the Grant Agreement GA101081953 attributed to the project H2OforAll—\emph{Innovative Integrated Tools and Technologies to Protect and Treat Drinking Water from Disinfection Byproducts (DBPs)} . Views and opinions expressed are, however, those of the author(s) only and do not necessarily reflect those of the European Union. Neither the European Union nor the granting authority can be held responsible for them. Funding for the work carried out by UK beneficiaries has been provided by UK Research and Innovation (UKRI) under the UK government’s Horizon Europe funding guarantee [grant number 10043071].
%}

\bibliographystyle{unsrt}
\bibliography{mybib}

\begin{thebibliography}{10}

\bibitem{berman2014geni}
Mark Berman, Jeffrey~S Chase, Lawrence Landweber, Akihiro Nakao, Max Ott,
  Dipankar Raychaudhuri, Robert Ricci, and Ivan Seskar.
\newblock Geni: A federated testbed for innovative network experiments.
\newblock {\em Computer Networks}, 61:5--23, 2014.

\bibitem{ertin2006kansei}
Emre Ertin, Anish Arora, Rajiv Ramnath, Vinayak Naik, Sandip Bapat, Vinod
  Kulathumani, Mukundan Sridharan, Hongwei Zhang, Hui Cao, and Mikhail
  Nesterenko.
\newblock Kansei: A testbed for sensing at scale.
\newblock In {\em Proceedings of the 5th international conference on
  Information processing in sensor networks}, pages 399--406, 2006.

\bibitem{keahey2020lessons}
Kate Keahey, Jason Anderson, Zhuo Zhen, Pierre Riteau, Paul Ruth, Dan
  Stanzione, Mert Cevik, Jacob Colleran, Haryadi~S Gunawi, Cody Hammock, et~al.
\newblock Lessons learned from the chameleon testbed.
\newblock In {\em Proceedings of the 2020 USENIX Conference on Usenix Annual
  Technical Conference}, pages 219--233, 2020.

\bibitem{carvalho2022can}
Gon{\c{c}}alo Carvalho, Filipe Magalh{\~a}es, Bruno Cabral, Vasco Pereira, and
  Jorge Bernardino.
\newblock Can we trust edge computing simulations? an experimental assessment.
\newblock {\em Computers}, 11(6):90, 2022.

\bibitem{svorobej2019simulating}
Sergej Svorobej, Patricia Takako~Endo, Malika Bendechache, Christos
  Filelis-Papadopoulos, Konstantinos~M Giannoutakis, George~A Gravvanis,
  Dimitrios Tzovaras, James Byrne, and Theo Lynn.
\newblock Simulating fog and edge computing scenarios: An overview and research
  challenges.
\newblock {\em Future Internet}, 11(3):55, 2019.

\bibitem{Nezami2021}
Zeinab Nezami, Kamran Zamanifar, Karim Djemame, and Evangelos Pournaras.
\newblock Decentralized edge-to-cloud load balancing: Service placement for the
  internet of things.
\newblock {\em IEEE Access}, 9:64983--65000, 2021.

\bibitem{behrisch2011sumo}
Michael Behrisch, Laura Bieker, Jakob Erdmann, and Daniel Krajzewicz.
\newblock Sumo--simulation of urban mobility: an overview.
\newblock In {\em Proceedings of SIMUL 2011, The Third International Conference
  on Advances in System Simulation}. ThinkMind, 2011.

\bibitem{gizinski2022design}
Tyler Gizinski and Xiang Cao.
\newblock Design, implementation and performance of an edge computing prototype
  using {R}aspberry {P}is.
\newblock In {\em 2022 IEEE 12th Annual Computing and Communication Workshop
  and Conference (CCWC)}, pages 0592--0601. IEEE, 2022.

\bibitem{pournaras2018decentralized}
Evangelos Pournaras, Peter Pilgerstorfer, and Thomas Asikis.
\newblock Decentralized collective learning for self-managed sharing economies.
\newblock {\em ACM Transactions on Autonomous and Adaptive Systems},
  13(2):1--33, 2018.

\bibitem{pournaras2020collective}
Evangelos Pournaras.
\newblock Collective learning: A 10-year odyssey to human-centered distributed
  intelligence.
\newblock In {\em 2020 IEEE International Conference on Autonomic Computing and
  Self-Organizing Systems (ACSOS)}, pages 205--214. IEEE, 2020.

\bibitem{vasisht2017farmbeats}
Deepak Vasisht, Zerina Kapetanovic, Jongho Won, Xinxin Jin, Ranveer Chandra,
  Sudipta Sinha, Ashish Kapoor, Madhusudhan Sudarshan, and Sean Stratman.
\newblock {F}arm{B}eats: An {I}o{T} platform for data-driven agriculture.
\newblock In {\em 14th USENIX Symposium on Networked Systems Design and
  Implementation (NSDI 17)}, pages 515--529, 2017.

\bibitem{zhang2019hetero}
Wuyang Zhang, Sugang Li, Luyang Liu, Zhenhua Jia, Yanyong Zhang, and Dipankar
  Raychaudhuri.
\newblock Hetero-edge: Orchestration of real-time vision applications on
  heterogeneous edge clouds.
\newblock In {\em Conf. on Computer Communications (INFOCOM)}, pages
  1270--1278. IEEE, 2019.

\bibitem{gedawy2016cumulus}
Hend Gedawy, Sannan Tariq, Abderrahmen Mtibaa, and Khaled Harras.
\newblock Cumulus: A distributed and flexible computing testbed for edge cloud
  computational offloading.
\newblock In {\em 2016 Cloudification of the Internet of Things (CIoT)}, pages
  1--6. IEEE, 2016.

\bibitem{hao2018edge}
Tianshu Hao, Yunyou Huang, Xu~Wen, Wanling Gao, Fan Zhang, Chen Zheng, Lei
  Wang, Hainan Ye, Kai Hwang, Zujie Ren, et~al.
\newblock Edge aibench: towards comprehensive end-to-end edge computing
  benchmarking.
\newblock In {\em International Symposium on Benchmarking, Measuring and
  Optimization}, pages 23--30. Springer, 2018.

\bibitem{meng2019dedas}
Jiaying Meng, Haisheng Tan, Chao Xu, Wanli Cao, Liuyan Liu, and Bojie Li.
\newblock Dedas: Online task dispatching and scheduling with bandwidth
  constraint in edge computing.
\newblock In {\em IEEE INFOCOM 2019-IEEE Conference on Computer
  Communications}, pages 2287--2295. IEEE, 2019.

\bibitem{munoz2017adrenaline}
Raul Mu{\~n}oz, Laia Nadal, Ramon Casellas, Michela~Svaluto Moreolo, Ricard
  Vilalta, Josep~Maria F{\`a}brega, Ricardo Mart{\'\i}nez, Arturo Mayoral, and
  Fco~Javier V{\'\i}lchez.
\newblock The adrenaline testbed: An {SDN/NFV} packet/optical transport network
  and edge/core cloud platform for end-to-end {5G} and {I}o{T} services.
\newblock In {\em 2017 European Conference on Networks and Communications
  (EuCNC)}, pages 1--5. IEEE, 2017.

\bibitem{pan2016homecloud}
Jianli Pan, Lin Ma, Ravishankar Ravindran, and Peyman TalebiFard.
\newblock Homecloud: An edge cloud framework and testbed for new application
  delivery.
\newblock In {\em 2016 23rd International Conference on Telecommunications
  (ICT)}, pages 1--6. IEEE, 2016.

\bibitem{royuela2022testbed}
Ignacio Royuela, Juan~Carlos Aguado, Ignacio de~Miguel, Noem{\'\i} Merayo,
  Ram{\'o}n J~Dur{\'a}n Barroso, Diego Hortelano, Lidia Ruiz, Patricia
  Fern{\'a}ndez, Rub{\'e}n~M Lorenzo, and Evaristo~J Abril.
\newblock A testbed for {CCAM} services supported by edge computing, and use
  case of computation offloading.
\newblock In {\em NOMS 2022-2022 IEEE/IFIP Network Operations and Management
  Symposium}, pages 1--6. IEEE, 2022.

\bibitem{rimal2018experimental}
Bhaskar~Prasad Rimal, Martin Maier, and Mahadev Satyanarayanan.
\newblock Experimental testbed for edge computing in fiber-wireless broadband
  access networks.
\newblock {\em IEEE Comm. Magazine}, 56(8):160--167, 2018.

\bibitem{xu2020support}
Qiaozhi Xu, Junxing Zhang, and Bulganmaa Togookhuu.
\newblock Support mobile fog computing test in {P}i{F}og{B}ed{II}.
\newblock {\em Sensors}, 20(7):1900, 2020.

\bibitem{xu2019pifogbed}
Qiaozhi Xu and Junxing Zhang.
\newblock {P}i{F}og{B}ed: a fog computing testbed based on {R}aspberry {P}i.
\newblock In {\em 2019 IEEE 38th International Performance Computing and
  Communications Conference (IPCCC)}, pages 1--8. IEEE, 2019.

\bibitem{boubin2022prowess}
Jayson Boubin, Avishek Banerjee, Jihoon Yun, Haiyang Qi, Yuting Fang, Steve
  Chang, Kannan Srinivasan, Rajiv Ramnath, and Anish Arora.
\newblock Prowess: An open testbed for programmable wireless edge systems.
\newblock In {\em Practice \& Experience in Advanced Research Computing}. 2022.

\bibitem{coutinho2018fogbed}
Antonio Coutinho, Fabiola Greve, Cassio Prazeres, and Joao Cardoso.
\newblock Fogbed: A rapid-prototyping emulation environment for fog computing.
\newblock In {\em International Conf. on Communications}, pages 1--7. IEEE,
  2018.

\bibitem{cappos2018edgenet}
Justin Cappos, Matthew Hemmings, Rick McGeer, Albert Rafetseder, and Glenn
  Ricart.
\newblock Edgenet: A global cloud that spreads by local action.
\newblock In {\em Symposium on Edge Computing (SEC)}, pages 359--360. IEEE,
  2018.

\bibitem{Gerostathopoulos2019}
Ilias Gerostathopoulos and Evangelos Pournaras.
\newblock Trapped in traffic? a self-adaptive framework for decentralized
  traffic optimization.
\newblock In {\em 14th International Symposium on Software Engineering for
  Adaptive and Self-Managing Systems}, pages 32--38. IEEE/ACM, 2019.

\bibitem{Davis2021}
Brionna Davis, Grace Jennings, Taylor Pothast, Ilias Gerostathopoulos,
  Evangelos Pournaras, and Raphael~E Stern.
\newblock Decentralized optimization of vehicle route planning—a cross-city
  comparative study.
\newblock {\em IEEE Internet Computing}, 25(3):34--42, 2021.

\bibitem{yfantis2021software}
Lampros Yfantis, Simon Stebbins, Ilias Gerostathopoulos, Tamara Djukic, Jordi
  Casas, David Garcia, Maria Kamargianni, and Manos Chaniotakis.
\newblock A software-agnostic agent-based platform for modelling emerging
  mobility systems.
\newblock In {\em 2021 7th International Conference on Models and Technologies
  for Intelligent Transportation Systems (MT-ITS)}, pages 1--6. IEEE, 2021.

\bibitem{slawomir2017next}
Gajewski Slawomir.
\newblock Next generation its implementation aspects in {5G} wireless
  communication network.
\newblock In {\em 2017 15th International Conference on ITS Telecommunications
  (ITST)}, pages 1--7. IEEE, 2017.

\bibitem{narayanan2022collective}
Arun Narayanan, Mohamed Korium, Dick~Carrillo Melgarejo, Hafiz~Majid Hussain,
  Arthur~Sousa De~Sena, Pedro Goria, Daniel Gutierrez-Rojas, Mehar Ullah, Ali
  Esmaeelnezhad, Mehdi Rasti, et~al.
\newblock Collective intelligence using {5G}: Concepts, applications, and
  challenges in sociotechnical environments.
\newblock {\em IEEE Access}, 2022.

\bibitem{etsi2018white}
Sami Kekki, Walter Featherstone, Yonggang Fang, Pekka Kuure, Alice Li, Anurag
  Ranjan, Debashish Purkayastha, Feng Jiangping, Danny Frydman, Gianluca Verin,
  et~al.
\newblock Mec in 5{G} networks.
\newblock {\em ETSI white paper}, 28(2018):1--28, 2018.

\end{thebibliography}

\end{document}